\documentclass[aps,pra,floatfix,twocolumn]{revtex4-1}
\usepackage[dvips]{graphicx}
\usepackage{color}

\usepackage{longtable}
\usepackage{dcolumn}
\usepackage[dvips]{graphicx}
\usepackage{bm}
\usepackage{bbm}

\usepackage{times}
\usepackage{nicefrac}
\usepackage{amsmath}
\usepackage{amsfonts}
\usepackage{amssymb}
\usepackage{amsthm}

\newcolumntype{w}[1]{D{.}{.}{#1}}

\newcolumntype{.}{D{x}{}{-1}}

\newcommand{\balpha}{\bm{\alpha}}
\newcommand{\bsigma}{\bm{\sigma}}
\newcommand{\bmu}{\bm{\mu}}
\newcommand{\bnabla}{\bm{\nabla}}

\newcommand{\vare}{\varepsilon}
\newcommand{\bfr}{{\bm {r}}}
\newcommand{\bfx}{{\bm {x}}}
\newcommand{\bfy}{{\bm {y}}}
\newcommand{\bfz}{{\bm {z}}}
\newcommand{\bfp}{{\bm {p}}}
\newcommand{\bfq}{{\bm {q}}}

\newcommand{\hr}{\hat{\bfr}}
\newcommand{\hx}{\hat{\bfx}}

\newcommand{\lbr}{\langle}
\newcommand{\rbr}{\rangle}

\newcommand{\pr}{^{\prime}}

\newcommand{\SixJ}[6]{
        \left\{
        \begin{array}{ccc}
        #1  & #2  & #3 \\
        #4  & #5  & #6 \\
        \end{array}
        \right\}
        }

\newcommand{\Za}{Z\alpha}

\begin{document}

\title{Theoretical energy levels of $\bm{1sns}$ and $\bm{1snp}$ states of helium-like ions}

\author{V.~A. Yerokhin}
\affiliation{Physikalisch-Technische Bundesanstalt, D-38116 Braunschweig, Germany}
\affiliation{Center for Advanced Studies, Peter the Great St.~Petersburg Polytechnic University,
195251 St.~Petersburg, Russia}

\author{A.~Surzhykov}
\affiliation{Physikalisch-Technische Bundesanstalt, D-38116 Braunschweig, Germany}
\affiliation{Technische Universit\"at Braunschweig, D-38106 Braunschweig, Germany}

\begin{abstract}

Energy levels of the $1sns$ and $1snp$ states of ions along the helium isoelectronic sequence
from carbon to uranium are calculated, with $n=3\,$--$\,7$. The computation is performed within
the relativistic configuration-interaction method, including the relativistic nuclear recoil
effect, the leading QED effects, and the frequency dependence of the Breit interaction. All
theoretical energies are supplied with uncertainty estimates.

\end{abstract}

\maketitle

\section*{Introduction}

Detailed knowledge of energy levels of various excited states of highly charged ions is required
for modelling spectra from high-temperature astrophysical sources, tokamak and laser-produced
plasmas \cite{chianti:19}. An analysis of spectra of these ions provides important information on
the electronic temperature and the ionization state of the plasma. Accurate values of wavelengths
of atomic transitions and their errors are required for interpretation of a wealth of
high-resolution data obtained in the last two decades by the Chandra X-ray Observatory and the
ESA’s X-ray Multi-Mirror Mission. The accuracy of theoretical results presently available is not
fully sufficient for this task \cite{smith:14}.

Helium-like ions are attractive objects for theoretical investigations. Because of their
simplicity, these ions can be described {\em ab initio} to very high accuracy, which is typically
better than what is accessible in present-day experiments. This makes theoretical transition
wavelengths of He-like ions suitable for usage as calibration references for experimental x-ray
spectra of multielectron ions \cite{stierhof:19}.

Among numerous calculations of He-like ions reported in the literature, most were concerned with
the ground and the first excited $n = 2$ states only. The most advanced of the calculations of the
$n \le 2$ states were performed within {\em ab initio} QED approaches in
Refs.~\cite{drake:88:cjp,artemyev:05:pra,yerokhin:10:helike,malyshev:19}. Higher excited states of
He-like ions attracted much less attention. Up to now, the most complete and reliable theoretical
energies of $n>2$ states were those obtained in 1985 by Vainstein and Safronova \cite{vainstein:85}
within the $1/Z$-expansion method (for $n \le 5$ and $Z \le 42$). The relative accuracy of their
energies was estimated to be not worse that $10^{-4}$, which was a significant achievement at the
time but is behind the state of the art nowadays.

The goal of the present work is to perform a large-scale calculation of energy levels of the $1sns$
and $1snp$ states of ions along the helium isoelectronic sequence. Since the ground and the $n = 2$
excited states are well covered in the literature, we will address the states with $n = 3\,$--$\,7$
in the present work. The summary of the previous theoretical results for the $n = 1$ and $n = 2$
states is given in supplementary material.

\section{Calculation}

Our present calculation closely follows the approach described in detail in our previous
investigations performed for Li-like ions \cite{yerokhin:17:lilike,yerokhin:18:lilike}. We use the
configuration-interaction (CI) method, in which the wave function of the atomic state $\Psi(PJM)$
with a definite parity $P$, the total angular momentum $J$, and the angular momentum projection $M$
is represented as a finite sum of the configuration-state functions (CSFs) with the same $P$, $J$,
and $M$,
\begin{equation}\label{eq4}
  \Psi(PJM) = \sum_r c_r \Phi(\gamma_r PJM)\,,
\end{equation}
where $\gamma_r$ denotes the set of additional quantum numbers that determine the CSF. The linear
coefficients $c_r$ in Eq.~(\ref{eq4}) and the energy of the atomic state $E$ are obtained by
solving the characteristic equation of the matrix of the Dirac-Coulomb-Breit (DCB) Hamiltonian
$H_{\rm DCB}$,
\begin{eqnarray}\label{eq:000}
    {\rm det} \Big(\lbr \gamma_r PJM|H_{\rm DCB}|\gamma_s PJM\rbr -E\,\delta_{rs}\Big) =
    0\,.
\end{eqnarray}

The energy levels obtained from the DCB Hamiltonian are supplemented by several corrections,
namely, the one-loop QED effects, the nuclear recoil effect, the frequency-dependence of the Breit
interaction, and higher-order QED contributions, as described in Ref.~\cite{yerokhin:18:lilike}.
The one-loop QED effects are accounted for by the model QED operator method
\cite{shabaev:13:qedmod}, as implemented by the QEDMOD package
\cite{shabaev:14:qedmod,shabaev:18:qedmod}.

\begin{table}
\caption{Comparison of present theoretical energies with previous theory by Vainstein and Safronova \cite{vainstein:85}, in Rydbergs.
\label{tab:compare:vain}
}
\begin{ruledtabular}
\begin{tabular}{lw{5.8}w{3.4}w{3.4}}
\multicolumn{1}{l}{State}
   & \multicolumn{1}{c}{Present}
      & \multicolumn{1}{c}{Vainstein \& Safronova}
           & \multicolumn{1}{c}{Difference}
\\
\hline\\[-5pt]
 \multicolumn{3}{l}{ Z =            6 } \\
$3^1\!S  $ &   25.98180\,(3) &          25.9839 &          -0.0021   \\
$3^3\!S  $ &   25.87625\,(3) &          25.8763 &           0.0001   \\
$3^3\!P_0$ &   25.98410\,(2) &          25.9820 &           0.0021   \\
$3^1\!P  $ &   26.05667\,(2) &          26.0568 &          -0.0001   \\
$4^1\!S  $ &   27.23086\,(3) &          27.2323 &          -0.0014   \\
$4^3\!S  $ &   27.18817\,(3) &          27.1877 &           0.0005   \\
$4^3\!P_0$ &   27.23214\,(2) &          27.2314 &           0.0007   \\
$4^1\!P  $ &   27.26217\,(2) &          27.2633 &          -0.0011   \\
$5^1\!S  $ &   27.80544\,(3) &          27.8064 &          -0.0010   \\
$5^3\!S  $ &   27.78404\,(3) &          27.7836 &           0.0004   \\
$5^3\!P_0$ &   27.80617\,(2) &          27.8064 &          -0.0002   \\
$5^1\!P  $ &   27.82139\,(2) &          27.8228 &          -0.0014   \\[5pt]

 \multicolumn{3}{l}{ Z =           14 } \\
$3^1\!S  $ &   160.1953\,(3) &         160.192 &           0.004   \\
$3^3\!S  $ &   159.8996\,(3) &         159.895 &           0.004   \\
$3^3\!P_0$ &   160.1848\,(1) &         160.179 &           0.006   \\
$3^1\!P  $ &   160.4162\,(1) &         160.412 &           0.004   \\
$4^1\!S  $ &   168.5189\,(3) &         168.516 &           0.003   \\
$4^3\!S   $ &   168.3991\,(3) &         168.388 &           0.011   \\
$4^3\!P_0$ &   168.5163\,(1) &         168.511 &           0.006   \\
$4^1\!P  $ &   168.6114\,(1) &         168.609 &           0.002   \\
$5^1\!S  $ &   172.3613\,(3) &         172.358 &           0.003   \\
$5^3\!S  $ &   172.3012\,(3) &         172.294 &           0.007   \\
$5^3\!P_0$ &   172.3605\,(1) &         172.355 &           0.005   \\
$5^1\!P  $ &   172.4086\,(1) &         172.405 &           0.003   \\[5pt]

 \multicolumn{3}{l}{ Z =           42 } \\
$3^1\!S  $ &   1558.502\,(4) &         1558.49 &            0.02   \\
$3^3\!S  $ &   1557.369\,(4) &         1557.36 &            0.00   \\
$3^3\!P_0$ &   1558.482\,(2) &         1558.47 &            0.02   \\
$3^1\!P  $ &   1561.880\,(2) &         1561.79 &            0.09   \\
$4^1\!S  $ &   1642.825\,(4) &         1642.81 &            0.01   \\
$4^3\!S  $ &   1642.366\,(4) &         1642.34 &            0.03   \\
$4^3\!P_0$ &   1642.822\,(2) &         1642.80 &            0.03   \\
$4^1\!P  $ &   1644.248\,(2) &         1644.20 &            0.05   \\
$5^1\!S  $ &   1681.658\,(4) &         1681.64 &            0.01   \\
$5^3\!S  $ &   1681.428\,(4) &         1681.41 &            0.02   \\
$5^3\!P_0$ &   1681.658\,(2) &         1681.63 &            0.02   \\
$5^1\!P  $ &   1682.385\,(2) &         1682.35 &            0.03   \\
\end{tabular}
\end{ruledtabular}
\end{table}
\begin{table}[t]
\caption{Comparison of theoretical and experimental energies of the $1sn\ell$ states in He-like ions
with $n = 3\,$--$\,7$, in Rydbergs.
\label{tab:compare:exp}
}
\begin{ruledtabular}
\begin{tabular}{lw{3.8}w{3.8}w{3.8}l}
\multicolumn{1}{l}{State}
   & \multicolumn{1}{c}{Theory}
      & \multicolumn{1}{c}{Experiment}
            & \multicolumn{1}{c}{Difference}
                & \multicolumn{1}{c}{Ref.}\\
\hline\\[-5pt]

 \multicolumn{3}{l}{ Z =            6 } \\
$3\,^3\!S  $ &   25.876\,25\,(3) &   25.876\,10\,(18) &    0.000\,14\,(18) &   \cite{engstrom:92} \\
$3\,^1\!P  $ &   26.056\,67\,(2) &   26.056\,48\,(19) &    0.000\,19\,(19) &   \cite{engstrom:92} \\
$3\,^3\!P  $ &   25.984\,29\,(2) &   25.984\,13\,(18) &    0.000\,17\,(18) &   \cite{engstrom:92} \\
$4\,^1\!S  $ &   27.230\,86\,(3) &   27.230\,90\,(24) &   -0.000\,04\,(24) &   \cite{engstrom:92} \\
$4\,^3\!S  $ &   27.188\,17\,(3) &   27.188\,06\,(18) &    0.000\,11\,(18) &   \cite{engstrom:92} \\
$4\,^1\!P  $ &   27.262\,17\,(2) &   27.262\,03\,(19) &    0.000\,14\,(19) &   \cite{engstrom:92} \\
$4\,^3\!P  $ &   27.232\,23\,(2) &   27.232\,07\,(18) &    0.000\,15\,(18) &   \cite{engstrom:92} \\
$5\,^1\!S  $ &   27.805\,44\,(3) &   27.805\,31\,(19) &    0.000\,13\,(19) &   \cite{engstrom:92} \\
$5\,^3\!S  $ &   27.784\,04\,(3) &   27.783\,92\,(18) &    0.000\,12\,(18) &   \cite{engstrom:92} \\
$5\,^1\!P  $ &   27.821\,39\,(2) &   27.821\,36\,(56) &    0.000\,03\,(56) &   \cite{engstrom:92} \\
$5\,^3\!P  $ &   27.806\,21\,(2) &   27.806\,06\,(18) &    0.000\,15\,(18) &   \cite{engstrom:92} \\
$6\,^3\!S  $ &   28.104\,11\,(3) &   28.104\,02\,(19) &    0.000\,09\,(19) &   \cite{engstrom:92} \\
$6\,^1\!P  $ &   28.125\,55\,(2) &   28.125\,70\,(91) &   -0.000\,15\,(91) &   \cite{engstrom:92} \\
$6\,^3\!P  $ &   28.116\,81\,(2) &   28.116\,68\,(18) &    0.000\,12\,(18) &   \cite{engstrom:92} \\
$7\,^3\!P  $ &   28.303\,60\,(2) &   28.303\,51\,(18) &    0.0000\,9\,(18) &   \cite{engstrom:92} \\[5pt]
 \multicolumn{3}{l}{ Z =            7 } \\
$3\,^1\!P  $ &   36.596\,40\,(2) &    36.596\,2\,(6) &     0.000\,2\,(6) &   \cite{engstrom:95} \\
$4\,^1\!P  $ &   38.334\,21\,(2) &    38.334\,6\,(8) &    -0.000\,4\,(8) &   \cite{engstrom:95} \\
$5\,^1\!P  $ &   39.140\,18\,(2) &     39.142\,(3) &     -0.002\,(3) &   \cite{engstrom:95} \\
$6\,^1\!P  $ &   39.578\,47\,(2) &     39.579\,(5) &     -0.001\,(5) &   \cite{engstrom:95} \\[5pt]
 \multicolumn{3}{l}{ Z =            8 } \\
$3\,^1\!P  $ &   48.918\,82\,(4) &     48.918\,(1) &      0.001\,(1) &   \cite{engstrom:95} \\
$4\,^1\!P  $ &   51.286\,32\,(3) &     51.286\,(2) &      0.000\,(2) &   \cite{engstrom:95} \\
$5\,^1\!P  $ &   52.384\,11\,(3) &     52.384\,(6) &      0.000\,(6) &   \cite{engstrom:95} \\[5pt]
 \multicolumn{3}{l}{ Z =           18 } \\
$3\,^1\!P  $ &   270.758\,0\,(2) &    270.766\,(9) &     -0.008\,(9) &   \cite{beiersdorfer:89} \\
$5\,^1\!P  $ &   291.299\,4\,(2) &     291.28\,(2) &       0.02\,(2) &   \cite{seely:85} \\[5pt]
 \multicolumn{3}{l}{ Z =           19 } \\
$3\,^1\!P  $ &   302.882\,9\,(3) &     302.89\,(1) &       0.00\,(1) &   \cite{seely:85} \\[5pt]
 \multicolumn{3}{l}{ Z =           21 } \\
$3\,^1\!P  $ &   372.618\,7\,(3) &     372.63\,(1) &      -0.01\,(1) &   \cite{beiersdorfer:89} \\
$4\,^1\!P  $ &   392.072\,3\,(3) &     392.08\,(1) &      -0.01\,(1) &   \cite{beiersdorfer:89} \\[5pt]
 \multicolumn{3}{l}{ Z =           22 } \\
$4\,^1\!P  $ &   431.696\,9\,(4) &     431.70\,(2) &       0.00\,(2) &   \cite{beiersdorfer:89} \\
$5\,^1\!P  $ &   441.632\,9\,(4) &     441.65\,(2) &      -0.02\,(2) &   \cite{beiersdorfer:89} \\[5pt]
 \multicolumn{3}{l}{ Z =           23 } \\
$3\,^1\!P  $ &   449.709\,4\,(4) &     449.73\,(3) &      -0.02\,(3) &   \cite{beiersdorfer:89} \\[5pt]
 \multicolumn{3}{l}{ Z =           24 } \\
$4\,^1\!P $ &   516.784\,8\,(5) &     516.76\,(2) &       0.02\,(2) &   \cite{beiersdorfer:89} \\
$5\,^1\!P $ &   528.711\,2\,(5) &     528.71\,(3) &       0.00\,(3) &   \cite{beiersdorfer:89} \\[5pt]
 \multicolumn{3}{l}{ Z =           26 } \\
$3\,^3\!P_1$ &   578.572\,2\,(6) &     578.60\,(1) &      -0.02\,(1) &   \cite{indelicato:86} \\
$3\,^1\!P $ &   579.256\,0\,(6) &     579.27\,(1) &      -0.02\,(1) &   \cite{indelicato:86} \\
$4\,^3\!P_1$ &   609.422\,7\,(6) &     609.44\,(2) &      -0.01\,(2) &   \cite{indelicato:86} \\
$4\,^1\!P $ &   609.706\,6\,(6) &     609.72\,(1) &      -0.02\,(1) &   \cite{indelicato:86} \\
            &                   &     609.69\,(2) &       0.02\,(2) &   \cite{beiersdorfer:89} \\
$5\,^1\!P $ &   623.806\,4\,(6) &     623.81\,(2) &       0.00\,(2) &   \cite{beiersdorfer:89} \\
\end{tabular}
\end{ruledtabular}
\end{table}

The uncertainty of the obtained theoretical energies comes from two main sources: the DCB energies
and the QED energy shifts, the QED uncertainty usually being the dominant one. For heavy ions, the
uncertainty due to experimental values of nuclear charge radii also contributes. The uncertainty of
the DCB energies was estimated by performing a series of CI calculations with 30-35 different basis
sets and by analysing consecutive increments of the results as the basis set was increased in
different directions. The uncertainty of the QED energy shifts was estimated by comparing the
results obtained by the QEDMOD package with those from rigorous QED calculations for the $n = 1$
and $n = 2$ states \cite{artemyev:05:pra,yerokhin:10:helike}. Basing on this comparison, we
estimate the uncertainty of the QEDMOD results as $(10/Z)\,$\% for the $1sns$ states and
$(5/Z)\,$\% for the $1snp$ states. For the fine-structure intervals, the QED corrections nearly
cancel, so we assume that the QED uncertainty is negligible.

With the help of the method outlined above we obtain the {\em total} energies of the excited $n \ge
3$ states of He-like ions. In order to obtain the {\em ionization} energies, we subtract from the
total energies delivered by the CI method the hydrogenic $1s$ energies from
Ref.~\cite{yerokhin:15:Hlike}. The case of $n = 2$ was used as an important cross-check against
more accurate full-scale QED calculations \cite{artemyev:05:pra,yerokhin:10:helike,malyshev:19}. In
order to obtain the energy differences of the excited-state levels and the ground-state $1s^2$
level, we use the ground-state energies obtained by compiling results of different {\em ab-initio}
QED calculations available in the literature, see supplementary material for details. For
completeness, the supplementary material also contains our compilation of theoretical energies of
the $(1s)^2$ and $1s2\ell$ states of He-like ions with the nuclear charge numbers from $Z = 6$ till
$Z = 92$, collected from different literature sources.

\section{Results}

Table~\ref{tab:compare:vain} presents a comparison of our theoretical energies with those
calculated by Vainstein and Safronova \cite{vainstein:85}. Energies relative to the ground $(1s)^2$
state are listed. We conclude that the results of Ref.~\cite{vainstein:85} are accurate typically
to 1-2 parts in $10^{-5}$ for ions around carbon and become even more accurate for heavier ions. At
the same time, we observe some systematic shifts for heavier ions, which are probably due to a
difference in the ground-state energies.

In Table~\ref{tab:compare:exp} we compare our theoretical energies with available experimental
results. Good agreement between theory and experiment is found nearly in all cases. Generally, we
conclude that the present theory is by one or two orders of magnitude more accurate than most of
the experimental data available to day.

Numerical results of our calculations of energy levels of the $1sns$ and $1snp$ states of He-like
ions are listed in Table~\ref{tab:en}. For each state, the ionization energy ($E_{\rm io}$) and the
energy relative to the ground $(1s)^2$ state ($E$) are supplied. For the $n\,^3\!P_J$ states, we
also provide the $(J = 1)$--$(J = 0)$ and $(J = 2)$--$(J = 1)$ fine-structure intervals ($E_{\rm
fs}$). For each ion, nuclear parameters used in calculations are specified. The nuclear charge
root-mean-square (rms) radii $R$ are taken from Ref.~\cite{angeli:13} and the nuclear masses $M$,
from Ref.~\cite{wang:12}. Theoretical energies are given with one or two uncertainties. When two
uncertainties are specified, the first one is the estimate of the theoretical error, whereas the
second one is due to the rms radius. When only one uncertainty is given, the second one is
negligible.

In summary, we have performed relativistic calculations of the energy levels of the $1sns$ and
$1snp$ excited states with $n=3\,$--$\,7$ for ions along the helium isoelectronic sequence. The
relativistic Dirac-Coulomb-Breit energies have been obtained by the configuration-interaction
method and supplemented with the QED energy shifts, relativistic recoil, and the
frequency-dependent shift of the Breit interaction. The results obtained are in good agreement with
previous theoretical and experimental data but significantly more accurate.

\section*{Acknowledgement}

V.A.Y. acknowledges support by the Ministry of Education and Science of the Russian Federation
Grant No.~3.5397.2017/6.7.


\begin{widetext}
\end{widetext}

\newpage

\begin{longtable}{lw{4.8}w{8.8}w{6.8}}
\caption{Theoretical energies of the $1sns$ and $1snp$ states of He-like ions, in Rydbergs.
 \label{tab:en}}
\\ \colrule\hline
\multicolumn{1}{c}{State}  &
                 \multicolumn{1}{c}{$E_{\rm io}$}  &
                             \multicolumn{1}{c}{$E$}  &
                                           \multicolumn{1}{c}{$E_{\rm fs}$}  \\
\hline            %
\\[-2pt]
\multicolumn{4}{l}{
  $Z = 6,\ \mbox{\rm $^{12}$C$^{4+}$},\  R = 2.4702\,(22)\ \mbox{\rm fm}, \  M/m = 21868.7$}
 \\[2pt]
$3\,^3\!S$                  &      2.941879\,(30) &     25.876245\,(31) &                     \\
$3\,^1\!S$                  &      2.836324\,(31) &     25.981800\,(31) &                     \\
$3\,^3\!P_0$                &      2.834021\,(15) &     25.984103\,(15) &                     \\
$3\,^3\!P_1$                &      2.834034\,(15) &     25.984091\,(15) &           -0.000012 \\
\end{longtable}

\newpage

\appendix

\section{Supplementary Material}

In this Supplementary Material we list theoretical energies of the $(1s)^2$ and $1s2\ell$ states of
He-like ions with the nuclear charge numbers from $Z = 6$ till $Z = 92$. Theoretical energies are
obtained by compiling results of different {\em ab-initio} QED calculations available in the
literature. Table~\ref{tab:en} lists theoretical energies, whereas Table~\ref{tab:compare} compares
results of different {\em ab-initio} QED calculations and those of the present compilation.

The tabulated theoretical energies are obtained as follows. For ions with $Z \le 12$, we take the
results of the nonrelativistic QED calculations by Yerokhin and Pachucki \cite{yerokhin:10:helike}.
Theoretical results for ions with $Z > 12$ are based on the calculation by Artemyev {\em et
al.}~\cite{artemyev:05:pra}, with several updates. First, the one-electron QED corrections for the
$1s$, $2s$, and $2p_j$ hydrogenic states used in Ref.~\cite{artemyev:05:pra} are updated according
to the review \cite{yerokhin:15:Hlike} and more recent studies of two-loop QED effects
\cite{czarnecki:16,yerokhin:18:sese}. Second, the results for the nuclear recoil effect calculated
by Malyshev {\em et al.}~\cite{malyshev:18} supersede results of a less accurate treatment in
Ref.~\cite{artemyev:05:pra}.

The comparison given in Table~\ref{tab:compare} demonstrates that for $Z = 12$, the results of two
independent QED calculations of Ref.~\cite{yerokhin:10:helike} and Ref.~\cite{artemyev:05:pra} are
in good agreement with each other. For larger $Z$, the calculation by Artemyev {\em et
al.}~\cite{artemyev:05:pra} is in good agreement with an independent computation by Malyshev {\em
et al.}~\cite{malyshev:19}.

\bibliographystyle{c:/-a-/papers/bibtex/phaip30}
\bibliography{c:/-a-/papers/bibtex/hfst}

\begin{widetext}
\end{widetext}

%
%
\begin{longtable}[t]{lw{7.8}w{7.8}}
\caption{Energy levels of the $n = 1$ and $n = 2$ states of He-like ions, in Rydbergs. $E_{\rm io}$
is the ionization energy, $E_{\rm tot}$ is the total binding energy, $R$ denotes the nuclear charge
root-mean-square (rms) radius, whereas $M/m$ is the ratio of the nuclear mass to the electron mass.
 \label{tab:en}}
\\ \colrule\hline
\multicolumn{1}{c}{State}  &
                 \multicolumn{1}{c}{$E_{\rm io}$}  &
                             \multicolumn{1}{c}{$E_{\rm tot}$}    \\
\hline            %
\\[-5pt]
\multicolumn{3}{l}{
  $Z = 6,\ \mbox{\rm $^{12}$C$^{4+}$},\  R = 2.4702\,(22)\ \mbox{\rm fm}, \  M/m = 21868.7$}
 \\[2pt]
%
%
%
$1\,^1\!S_0$    &    28.8181243\,(29) &    64.8319608\,(29) \\
$2\,^1\!S_0$    &    6.44603603\,(28) &   42.45987254\,(28) \\
$2\,^3\!S_1$    &    6.84480754\,(31) &   42.85864405\,(31) \\
$2\,^3\!P_0$    &   6.444772362\,(74) &  42.458608874\,(75) \\
\end{longtable}

\begin{table*}
\caption{Comparison of different theoretical results for ionzation
energies of the $n=1$ and $n = 2$ states in He-like ions, in eV. 1~hartree = 2~Rydberg = 27.211\,386\,245\,988~eV.
\label{tab:compare}
}
\begin{ruledtabular}
\begin{tabular}{lw{3.8}w{3.8}w{3.8}w{3.8}}
 &
\multicolumn{1}{l}{Present compilation}
   & \multicolumn{1}{c}{Yerokhin\&Pachucki \cite{yerokhin:10:helike}}
      & \multicolumn{1}{c}{Artemyev et al. \cite{artemyev:05:pra}}
                & \multicolumn{1}{c}{Malyshev et al. \cite{malyshev:19}}\\
\hline\\[-5pt]
\multicolumn{3}{l}{
  $Z = 12$ }
 \\[2pt]
%
%
%
$1\,^1\!S_0$    &     1761.8049\,(10) & 1761.8049\,(10)  & 1761.8047\,(2) & \\
$2\,^1\!S_0$    &      417.9686\,(1)  &  417.9686\,(1)   &  417.9689\,(1) & \\
$2\,^3\!S_1$    &      430.6929\,(1)  &  430.6929\,(1)   &  430.6929\,(1) & \\
$2\,^3\!P_0$    &      418.80914\,(2) &  418.80914\,(2)  &  418.8092\,(1) & \\
$2\,^3\!P_1$    &      418.70596\,(2) &  418.70596\,(2)  &  418.7059\,(1) & \\
\end{tabular}
\end{ruledtabular}
\end{table*}

\end{document}